\documentstyle[12pt]{article}
\newcommand{\bra}{\langle}
\newcommand{\ket}{\rangle}
\newcommand{\R}{\mbox{\boldmath $ R $}}
\newcommand{\Z}{\mbox{\boldmath $ Z $}}
\newcommand{\C}{\mbox{\boldmath $ C $}}

\begin{document}
\baselineskip 6mm
\begin{flushright}
KU-AMP 95010 \\
INS-Rep.-1108 \\
hep-th/9508165 \\
August 1995
\end{flushright}
\vspace*{5mm}
\begin{center}
{\Large\bf
Induced Gauge Fields in the Path Integral}
\vspace{15mm} \\
{\sc Shogo TANIMURA} \\
{\it Department of Applied Mathematics and Physics \\
Faculty of Engineering,
Kyoto University \\
Kyoto 606-01,
Japan}\footnote{%
E-mail: tanimura@kuamp.kyoto-u.ac.jp}
\vspace{10mm} \\
{\sc Izumi TSUTSUI} \\
{\it Institute for Nuclear Study,
University of Tokyo \\
Midori-cho, Tanashi-shi, Tokyo 188,
Japan}\footnote{%
E-mail: tsutsui@ins.u-tokyo.ac.jp}
\vspace{20mm} \\
\center{\bf Abstract} \\

\begin{minipage}[t]{120mm}
\baselineskip 5mm
{\small
The path integral on a homogeneous space $G/H$ is constructed,
based on the guiding principle
`first lift to $G$ and then
project to $G/H$'.  It is then shown that
this principle admits inequivalent quantizations inducing
a gauge field (the canonical connection) on the homogeneous
space, and thereby reproduces the result
obtained earlier by algebraic approaches.
}

\end{minipage}
\end{center}

\newpage
\baselineskip 16pt
\parskip 6pt
\section{Introduction}
Geometric approaches to quantum mechanics have
been studied by various groups ever
since the foundation of quantum mechanics
was laid down.
The prime aim of such approaches is
to render quantum mechanics applicable to more general settings, not just
to Euclidean space as originally done.
However, it is by now well recognized that quantization is
generally difficult to carry out unless
the setting is fairly simple.  A system whose classical configuration
space $Q$ is a homogeneous space given by a coset
$G/H$ falls into this simple category.
An important lesson learned when quantizing on
homogeneous spaces
is that
there are actually (infinitely) many
{\it inequivalent quantizations}
allowed~\cite{Mackey,Doebner,Isham}.
In other words, there exist
unitarily inequivalent Hilbert spaces where
physical properties, such as their energy spectra, may differ
from each other.
These inequivalent quantizations are classified
according to the induced representation~\cite{Mackey} which is used
for the quantization.
\par
Interest in the inequivalent quantizations has been renewed recently
after Landsman and Linden examined the
physical implications of the quantizations
and found that a special type of
gauge field is induced
on homogeneous
spaces~\cite{Landsman,LL,Ohnuki,McMullan,Robson}.
The gauge fields are a (topological) solution
of the Yang-Mills equation on the spaces,
called the {\it canonical connection} (or $H$-{\it connection}).
However, the previous arguments leading to the gauge fields
are algebraic and abstract, and there is no intuitive
account of this rather mysterious appearance of gauge fields.
It would be therefore desirable
to develop a path integral account,
which normally admits a more intuitive understanding
based on the geometry of the configuration space.
In this note we wish to take a step in this direction ---
we shall show that,
the path integral on a homogeneous space carries
the canonical connection as a gauge field, if we adopt
the guiding principle that the path integral be constructed first on
the group manifold $G$ and then projected down to the homogeneous space
$G/H$.  This \lq first lift and then project' principle may be
arguable, but it is certainly true that the case $Q = S^1$, where
the path integral is known to reproduce the inequivalent
quantizations correctly~\cite{Schulman}, relies on this principle.
We shall not dwell on this issue until the end of the paper
where a possible explanation is given.
We here mention that similar induced gauge fields appear in various
other contexts as well; {\it e.g.},
in the context of Berry's (geometric) phase
in quantum mechanics~\cite{Berry,Simon,Wilczek,Levay}
or in the kinematics of molecules and deformable
bodies~\cite{Guichardet,Iwai,Shapere,Montgomery}.
Moreover, induced gauge fields
play an important role in high energy physics too; {\it e.g.},
in the so-called hidden local symmetry
of nonlinear sigma models~\cite{Bando}
and in the search for
a possible origin of dynamical gauge bosons~\cite{Kikkawa,Tamura}.
We hope that the path integral account given in this paper
may shed some light on the
machinery for those phenomena in general.
\par
The plan of this paper is as follows.
First we review quantum mechanics on $Q = S^1 $
to see how the gauge field is induced.
Motivated by this simple example, we then generalize the construction
of the path integral
to the case of homogeneous spaces $ Q = G/H $ following the
above guiding principle.
We shall find that the canonical connection
does appear in the path integral in the form expected, once
the induced representation is incorporated in the path integral scheme.
Finally we will argue a possible generalization
to inhomogeneous spaces, together with a restriction
that may underlie the guiding principle we adopted.
%
%
\section{Covering the path integral}
Let us begin by reviewing
the path integral on a circle $ S^1 $ \cite{Schulman}.
(A further discussion can be found in refs.\cite{LL,S1}.)
We first regard $ S^1 $ as the coset
$S^1 \cong \R / 2 \pi \Z $ by identifying
the point $ x $ of $ \R $ with other points $ x + 2 \pi n$ for
$ n \in \Z$.
This identification defines a covering map $ \pi : \R \to S^1 $.
Our idea is then to construct the path integral on $S^1$ from the
path integral on $\R$ with the above identification in mind.
\par
Let $ K_{R} ( x', x ; t ) = \bra x' | e^{ -i H t } | x \ket $
be a propagator on $ \R $ which is
invariant under the translation by
$2\pi$,
\begin{equation}
	K_{R} ( x'+ 2 \pi , x + 2 \pi ; t )
	=
	K_{R} ( x' , x ; t ).
	\label{0}
\end{equation}
On account of the
identification of points $ x' + 2 \pi n $ with $ x' $,
summation over $ n $ may lead to a propagator on $ S^1 $;
\begin{equation}
	K_{S^1} ( x', x ; t )
	=
	\sum_{ n = - \infty }^{ \infty }
	K_{R} ( x' + 2 \pi n , x ; t ),
	\label{1}
\end{equation}
where we interpret the integer $ n $ as the winding number
of a path connecting two points $ x $ and $ x' $ along the circle $ S^1 $.
Clearly, this expression admits an immediate generalization.
In fact, we do not have an {\it a priori} physical
reason to add up propagators for different winding numbers
with the same weight, as long as the weight is a phase
factor.  Based on this observation
Schulman~\cite{Schulman} proposed to insert a weight factor $ \omega_n $
with $ | \omega_n | = 1 $ to obtain a more general propagator
\begin{equation}
	K_{S^1}^{\omega} ( x' , x ; t )
	=
	\sum_{ n = - \infty }^{ \infty } \,
	\omega_n
	K_{R} ( x' + 2 \pi n , x ; t ).
	\label{2}
\end{equation}
The composition law of the propagator
\begin{equation}
	\int_{0}^{2 \pi} dx' \,
	K_{S^1}^{\omega} ( x'', x' ; t' ) \,
	K_{S^1}^{\omega} ( x' , x  ; t  )
	=
	K_{S^1}^{\omega} ( x'', x  ; t + t' )
	\label{3}
\end{equation}
is guaranteed if the weight satisfies\footnote{%
The weight can actually be determined by requiring consistency against
a shift of winding numbers~\cite{Schulman} but here we use the composition
law for our later generalization.}
\begin{equation}
	\omega_m \, \omega_n = \omega_{m+n}.
	\label{4}
\end{equation}
This implies that $ \omega : \pi_1 ( S^1 ) \to U(1) $ is
a unitary representation of the first homotopy group
$ \pi_1 ( S^1 ) \cong \Z $ and hence
given by $ \omega_n = e^{ i \alpha n } $
with a real parameter $ \alpha \in [ 0 , \, 2 \pi ) $.
For each value of $ \alpha $,
the propagator (\ref{2}) furnishes
an inequivalent quantum theory on $ S^1 $.
To see the physical meaning of $ \alpha $,
we assume $ K_{R} $ to be of the standard form
\begin{equation}
	K_{R} ( x' , x ; t )
	=
	\int_x^{x'} [dx] \,
	\exp
	\left[ i
		\int dt
		\biggl\{
			\frac{1}{2} \biggl( \frac{ d x }{ d t } \biggr)^2
			- V ( x )
		\biggr\}
	\right],
	\label{5}
\end{equation}
where $ V( x + 2 \pi ) = V( x ) $ in order to satisfy (\ref{0}).
Putting $ A = \alpha / ( 2 \pi ) $, we find that the propagator
(\ref{2}) can be rewritten as
\begin{equation}
	K_{S^1}^{\omega} ( x' , x ; t )
	=
        e^{-i\frac{\alpha}{2\pi}( x' - x )}
	\sum_{ n = - \infty }^{ \infty } \,
	\int_x^{ x' + 2 \pi n } [ dx ] \,
	\exp
	\left[ i
		\int dt
		\biggl\{
			\frac{1}{2} \biggl( \frac{ d x }{ d t } \biggr)^2
			- V ( x )
			+ A \, \frac{ d x }{ d t }
		\biggr\}
	\right].
	\label{6}
\end{equation}
We therefore see that the insertion
of the weight $ \omega_n = e^{ i \alpha n } $
just amounts to
introduction of the minimal coupling with the vector potential $ A $.
Being constant, the vector potential has vanishing curvature on
$S^1$ but the flux penetrating the circle is finite.  Hence, its
physical consequence is analogous to that of the Aharonov-Bohm effect.
%
%
\section{Lifting the path integral}
What we have considered above is a covering
$ \pi : \R \to S^1 \cong \R / 2 \pi \Z $.
A point $ x' $ in $ S^1 $ is lifted to points $ x' + 2 \pi n $ in $ \R $,
which are translated by the action of the group $ \Z $.
For each lifted point a propagator in $ \R $ is defined,
then we add them up
with a weight factor given by the representation $ \omega : \Z \to U(1) $
to obtain a propagator in $ S^1 $.
Thus a path in $ S^1 $ is lifted up to $ \R $ once,
and then it is projected down to $ S^1 $ with a nontrivial weight
multiplied, resulting in inequivalent quantizations and inducing a
$ U(1) $ gauge field.
In this section, we shall repeat the above construction
of the path integral to a homogeneous space $G/H$, where
$ G $ is a compact Lie group and $ H $ its closed subgroup.
In order to set up a framework where
a generalization of the covering $ \R \to S^1 \cong \R / 2 \pi \Z $
can be realized for $Q \cong G/H$,
we take the principal fiber bundle $ \pi : G \to G/H $
in which $ H $ acts on $ G $ from the right
and $ G $ acts on $ G/H $ from the left.
The difference from the former case is that
$ H $ can be a continuous group or a nonabelian group in general, and
hence
the summation over the winding numbers $ \sum_n \, ( n \in \Z ) $
will be replaced by the integration over the group $ \int_H \, d h $.
For a nonabelian $ H $
its 1-dimensional representation is always trivial, but
if we use higher dimensional nontrivial representations
we will get inequivalent quantizations, as we shall see below.
\par
According to our guiding principle,
we first lift our system from $Q$ to $ G $, and consider
a propagator in $ G $ which is a map
$ K_G : G \times G \times \R^+ \to \C $.  The
propagator we are interested in is one which is invariant under the
$H$ action (as in (\ref{0})),
\begin{equation}
	K_G ( g'h , g h ; t ) = K_G ( g' , g ; t )\ ,
	\label{7}
\end{equation}
for arbitrary $ g, \, g' \in G $ and $ h \in H $.
As before, we take the standard form for the propagator
on $ G $,
\begin{equation}
	K_G ( g' , g ; t )
	=
	\int_g^{g'} [dg] \,
	\exp
	\left[ i
		\int dt
		\biggl\{
			\frac{1}{2}
			\biggl| \biggl| \frac{ d g }{ d t } \biggr| \biggr|^2
			- V ( g )
		\biggr\}
	\right]\ .
	\label{8}
\end{equation}
Since the condition (\ref{7}) implies
the invariance of the potential
$ V( g h ) = V( g ) $,
which corresponds to the
periodicity $ V( x + 2 \pi ) = V( x ) $ in (\ref{5}),
the potential $ V $ is actually a function of the homogeneous space
$ V : Q \to \R $.
The norm $ || \cdot || $ used in (\ref{8}) is given
by the invariant metric on $ G $, that is,
$ || \dot{g} ||^2 = \mbox{Tr} ( g^{-1} \dot{g} )^2 $ where
\lq $\mbox{Tr}$' is a matrix trace properly normalized in
some irreducible representation.
(The expression (\ref{8}) is rather symbolic; for
a concrete expression, see \cite{LL}.)
Now we define two unitary operators $ U_t $ and $ R_h $
acting on $ \psi \in L_2 ( G ) $ by
\begin{eqnarray}
	&&
	( U_t \psi ) ( g' )
	=
	\int_G dg \, K_G ( g' , g ; t ) \, \psi ( g )\ ,
	\label{10}
	\\
	&&
	( R_h \psi ) ( g )
	=
	\psi ( g h )\ ,
	\label{11}
\end{eqnarray}
for each $ t > 0 $ and $ h \in H $, where
$ dg $ in (\ref{10}) is the normalized Haar measure of $ G $.
Then,
the invariance (\ref{7}) states that $ U_t \, R_h = R_h \, U_t $,
and hence
there exists a conserved quantity associated with this invariance.
Consequently, the Hilbert space $ L_2(G) $ can be decomposed
into the irreducible representations of $ H $.
\par
To implement the decomposition,
let $ ( V_\chi , \rho_\chi ) $
be an irreducible unitary representation of $ H $,
where $ V_\chi $ is a representation space
labeled by $ \chi $.
A function $ f : G \to V_\chi $ is called
{\it $ \chi $-equivariant}
if it satisfies $ f ( g h ) = \rho_\chi( h )^{-1} \, f ( g ) $.
In other words, $ f $ is a section
of the associated vector bundle $ E_\chi = G \times_\rho \, V_\chi $.
The space of $ \chi $-equivariant functions is denoted by $ \Gamma^\chi $,
which is equipped with the inner product
\begin{equation}
	\bra f_1 , f_2 \ket
	=
	\int_G dg \, \bra f_1 ( g ) , f_2 ( g ) \ket,
	\label{12}
\end{equation}
where in the right-hand side $ \bra \cdot , \cdot \ket $
denotes the inner product of the linear space $ V_\chi $.
Consider then the operator
$ I^{ ( \chi , j ) } : L_2(G) \to \Gamma^\chi $ defined by
\begin{equation}
	( I^{ ( \chi , j ) } \psi )^i ( g )
	=
	\sqrt{ d_\chi }
	\int_H dh \,
	\rho_{\chi}^{ ij } ( h ) \, \psi ( g h )\ .
	\label{13}
\end{equation}
Here $ d_\chi = \dim V_\chi $,
the indices $ i, j = 1, \cdots , d_\chi $ run over the
components of $ V_\chi $,
$ \rho_{\chi}^{ ij } ( h ) $ is a matrix element of
an unitary representation
$ \rho_{\chi} ( h ) $, and
$ dh $ the normalized Haar measure of $ H $.
This operator $ I^{ ( \chi , j ) } $ provides a partial isometry
in the sense that
$ I^{ ( \chi , j ) } $ is isometric on
$ ( \ker I^{ ( \chi , j ) } )^{\perp} $
(for more detail on $ I^{ ( \chi , j ) } $, see \cite{LL}).
The adjoint operator
$ I^{ ( \chi , j ) \dagger } : \Gamma^\chi \to L_2(G) $
is defined by the relation
$ \bra I^{ ( \chi , j ) \dagger } f , \psi \ket =
  \bra f , I^{ ( \chi , j ) } \psi \ket $,
where the former bracket is the inner product of $ L_2(G) $
while the latter is the one of $ \Gamma^{ \chi } $.
One can then show that
$ I^{ ( \chi , j ) \dagger } $ picks up the $j$-th component
of a $\chi$-equivariant function:
\begin{equation}
	( I^{ ( \chi , j ) \dagger } f ) ( g )
	=
	\sqrt{ d_\chi }
	f^j ( g ).
	\label{13.1}
\end{equation}
\par
Next let us turn to the time evolution operator $ U_t $.
Observe first that, thanks to the invariance (\ref{7}), the product
$ U_t^{ ( \chi , j ) } =
I^{ ( \chi , j ) } \, U_t \, I^{ ( \chi , j ) \dagger } $
may be used to
define a unitary time evolution projected on $ \Gamma^\chi $.
Explicitly, it is given by
\begin{equation}
	( U_t^{ ( \chi , j ) } f )^i ( g' )
	=
	\int_G dg
	\int_H dh
	\sum_{ k=1 }^{ d_\chi } \,
	\rho_\chi ( h )^{ ik } \, K_G ( g' h , g ; t ) f^k ( g ),
	\label{13.2}
\end{equation}
which shows that $ U_t^{ ( \chi , j ) } $ is in fact
independent of $ j $, and hence
can be written simply as $ U_t^\chi $.
 From this expression
we can deduce the projected propagator $ K_Q^\chi $
acting on $ \Gamma^\chi $ via
$( U_t^\chi f ) ( g' )=
	\int_G dg \, K_Q^\chi ( g' , g ; t ) \, f ( g )$,
that is,
\begin{equation}
        K_Q^\chi ( g' , g ; t )
        =
        \int_H dh \, \rho_\chi ( h ) \, K_G ( g' h , g ; t )\ .
        \label{14}
\end{equation}
The projected propagator $ K_Q^\chi $ is a map
$ K_Q^\chi : G \times G \times \R^+ \to \mbox{End} ( V_\chi ) $,
which is an analogue of (\ref{2}).
Note that
the summation $ \sum_n \, ( n \in \Z ) $ with respect to covering points
is replaced by the integration $ \int_H dh $ along the
fiber as planned, whereas the phase factor $ \omega_n $ is now replaced
by the nonabelian weight $ \rho_\chi (h) $.
Note also that the composition law
$ U_{t+t'}^\chi = U_{t'}^\chi U_t^\chi  $
is ensured by the
homomorphism $ \rho_\chi (h' h) = \rho_\chi (h') \rho_\chi (h) $
of the representation.
The projected propagator $ K_Q^\chi $ has the following properties,
\begin{eqnarray}
	&&
	K_Q^\chi ( g' h , g ; t )
	=
	\rho_\chi ( h )^{ -1 } \, K_Q^\chi ( g' , g ; t )\ ,
	\label{16}
	\\
	&&
	K_Q^\chi ( g' , g h ; t )
	=
	K_Q^\chi ( g' , g ; t ) \, \rho_\chi ( h )\ .
	\label{17}
\end{eqnarray}
Thus we see that our path integral on $ Q $ has successfully
accommodated the
inequivalent quantizations which are labeled by
the irreducible representation $ \chi $ of $H$.
%
%
\section{Inducing the gauge field}
Having found the path integral which reproduces the
inequivalent quantizations obtained in algebraic approaches,
we now move on to examine whether it
carries the canonical connection
as a gauge field in the form of the (nonabelian) minimal coupling,
as we have seen
in (\ref{6}) for the case $S^1$.
This requires to analyze the local structure
of the propagator (\ref{16}) by dividing a path in $ Q $ into small intervals,
and for this we need some preparations.
\par
Recall first that
the Haar measure $ dg $ of $ G $
induces the $ G $-invariant measure $ dq = \pi_\ast ( dg ) $ on $ Q $,
whereby a function $ \phi : Q \to \C $ can be integrated as
\begin{equation}
	\int_Q dq \, \phi( q )
	=
	\int_G dg \, \phi( \pi ( g ) ).
	\label{18}
\end{equation}
Let $ \{ D_\alpha \} $ be an open covering of
$ Q = \cup_\alpha D_\alpha $,
$ \{ s_\alpha : D_\alpha \to G \} $ be a set of local sections
of the fiber bundle $ \pi : G \to Q $, and
$ \{ w_\alpha : Q \to \R \} $ be a partition of unity
associated with the covering $ \{ D_\alpha \} $.
Then they give local expressions to various objects:
for a $ \chi $-equivariant function $ f $
its pullback is
$ f_\alpha = s_\alpha^\ast f = f \circ s_\alpha : D_\alpha \to V_\chi $;
the pullback of the projected propagator $ K_Q^\chi $ is a map
$ K_{ \alpha \beta }^{ \chi } :
D_\alpha \times D_\beta \times \R^+ \to \mbox{End} ( V_\chi ) $
defined by $ K_{ \alpha \beta }^{ \chi } ( q' , q ; t )
= K_Q^\chi ( s_\alpha( q' ), s_\beta( q ) ; t ) $; and
if $ q' \in D_\alpha \cap D_\gamma $
the local expressions are related by
$ K_{ \gamma \beta }^\chi ( q' , q ; t )
= \rho_\chi ( t_{ \gamma \alpha } ( q' ) )
K_{ \alpha \beta }^\chi ( q' , q ; t ) $
with a transition function
$ t_{ \gamma \alpha } ( q' ) = s_\gamma ( q' )^{-1} s_\alpha ( q' ) $.
In terms of these, the time evolution operator reads
\begin{equation}
	( U_t^\chi f )_\alpha ( q' )
	=
	\sum_\beta
	\int_Q dq \, K_{ \alpha \beta }^\chi ( q' , q ; t ) \,
	w_\beta ( q ) f_\beta ( q ).
	\label{19}
\end{equation}
Hence the composition law
$ U_{ t + t' }^\chi = U_{ t' }^\chi U_{ t }^\chi $
implies
\begin{eqnarray}
	K_Q^\chi ( g'', g  ; t + t' )
	& = &
	\int_G dg' \,
	K_Q^\chi ( g'', g' ; t' ) \,
	K_Q^\chi ( g' , g  ; t  )
	\nonumber
	\\
	& = &
	\sum_\alpha
	\int_{ D_\alpha } dq' \,
	w_\alpha (q') \,
	K_Q^\chi ( g'', s_\alpha (q') ; t' ) \,
	K_Q^\chi ( s_\alpha (q') , q  ; t  ).
	\label{20}
\end{eqnarray}
Inserting intermediate points repeatedly, we obtain
\begin{eqnarray}
	K_Q^\chi ( g_n , g_0  ; t )
	& = &
	\sum_{ \alpha_1 , \cdots , \alpha_{ n-1 } }
	\int_{ D_{\alpha_{n-1}} } dq_{n-1}
	\cdots
	\int_{ D_{\alpha_1} } dq_1 \,
	w_{ \alpha_{n-1} } ( q_{n-1} ) \cdots w_{ \alpha_1 } ( q_1 )
	\nonumber
	\\
	&&
	\qquad \times\,
	K_Q^\chi ( g_{ n },
	           s_{ \alpha_{ n-1 } } ( q_{ n-1 } ) ; \epsilon )\,
	K_Q^\chi ( s_{ \alpha_{ n-1 } } ( q_{ n-1 } ) ,
	           s_{ \alpha_{ n-2 } } ( q_{ n-2 } ) ; \epsilon )
	\cdots
	\nonumber
	\\
	&&
	\qquad \times\,
	K_Q^\chi ( s_{ \alpha_{ 2 } } ( q_{ 2 } ) ,
	           s_{ \alpha_{ 1 } } ( q_{ 1 } ) ; \epsilon )\,
	K_Q^\chi ( s_{ \alpha_{ 1 } } ( q_{ 1 } ) ,
	           g_{ 0 } ; \epsilon ),
	\label{21}
\end{eqnarray}
where $ \epsilon = t / n $.
When two points
$ q_k = q( \tau ) $ and $ q_{ k+1 } = q ( \tau + \epsilon ) $
are close enough to
be contained in a single patch $ D_\alpha $,
one of the factorized propagator becomes
\begin{equation}
	K_{ \alpha \alpha }^\chi
	( q ( \tau + \epsilon ) , q ( \tau ) ; \epsilon )
	=
	\int_H dh \, \rho_\chi ( h ( \epsilon ) ) \,
	K_G ( s_\alpha ( q ( \tau + \epsilon ) ) h ( \epsilon ) ,
	      s_\alpha ( q ( \tau            ) )   ; \epsilon ),
	\label{22}
\end{equation}
where we extend $ h \in H $ to be a smooth function
$ h : ( - \epsilon , \epsilon ) \to H $ such that $ h(0) = e $
($e$ is the identity element of $H$) and $ h( \epsilon ) = h $.
Then eq.(\ref{8}) tells that for a short time interval $ \epsilon $,
\begin{eqnarray}
	&&
	K_G ( s_\alpha ( q ( \tau + \epsilon ) ) h ( \epsilon ) ,
	      s_\alpha ( q ( \tau            ) )   ; \epsilon )
	\nonumber \\
	&&
	\qquad \quad
	\approx
	\exp
	\left[
		i \epsilon
		\biggl\{
			\frac{1}{2}
			\biggl| \biggl|
				\Bigl.
					\frac{d}{ d \epsilon }
					s_\alpha ( q ( \tau + \epsilon ) )
                                        h ( \epsilon )
				\Bigr|_{ \epsilon = 0 }
			\biggr| \biggr|^2
			- V ( q ( \tau ) )
		\biggr\}
	\right],
	\label{23}
\end{eqnarray}
with
\begin{eqnarray}
	&&
	\biggl| \biggl|
		\Bigl.
			\frac{d}{ d \epsilon }
			s_\alpha ( q ( \tau + \epsilon ) ) h ( \epsilon )
		\Bigr|_{ \epsilon = 0 }
	\biggr| \biggr|^2
	\nonumber
	\\
	& = &
	\mbox{Tr}
	\biggl[
		\Bigl.
			h(\epsilon)^{-1}
			s_\alpha ( q ( \tau ) )^{ -1 }
			\frac{ d s_\alpha ( q ( \tau ) ) }{ d \tau }
			h(\epsilon)
		\Bigr|_{ \epsilon = 0}
		+
		\Bigl.
			h(\epsilon)^{-1}
			\frac{ d h ( \epsilon ) }{ d \epsilon }
		\Bigr|_{ \epsilon = 0}
	\biggr]^2
	\nonumber
	\\
	& = &
	\mbox{Tr}
	\biggl[
		s_\alpha ( q ( \tau ) )^{ -1 }
		\frac{ d s_\alpha ( q ( \tau ) ) }{ d \tau }
		+
		\Bigl.
			\frac{ d h ( \epsilon ) }{ d \epsilon }
			h(\epsilon)^{-1}
		\Bigr|_{ \epsilon = 0 }
	\biggr]^2
	\nonumber
	\\
	& = &
    \mbox{Tr}
    \biggl[
		P_{ \cal H }
		\Bigl(
			s_\alpha ( q ( \tau ) )^{ -1 }
			\frac{ d s_\alpha ( q ( \tau ) ) }{ d \tau }
		\Bigr)
       	+
		\Bigl.
        	\frac{ d h ( \epsilon ) }{ d \epsilon }
			h(\epsilon)^{-1}
		\Bigr|_{ \epsilon = 0 }
	\biggr]^2
	\nonumber
	\\
	&&
	+
	\,
	\mbox{Tr}
	\biggl[
		P_{ \cal H }^{ \perp }
		\Bigl(
			s_\alpha ( q ( \tau ) )^{ -1 }
			\frac{ d s_\alpha ( q ( \tau ) ) }{ d \tau }
		\Bigr)
	\biggr]^2,
	\label{24}
\end{eqnarray}
where
$ P_{\cal H} $ is a projector from the Lie algebra
of $ G $ onto the Lie algebra of $ H $,
and $ P_{ \cal H }^{ \perp } $ denotes its orthogonal complement.
\par
Now, if we take the interval $\epsilon$ small enough, then
the contribution from the stationary point of
(\ref{24}) with respect to the variation of $ h $ will dominate
in the integration $ \int_H dh $ in (\ref{22}).
Thus in the limit $\epsilon \rightarrow 0$
the integration may be replaced by the value at the stationary point
\begin{equation}
	\Bigr.
		\frac{ d h ( \epsilon ) }{ d \epsilon }
		h(\epsilon)^{-1}
	\Bigr|_{ \epsilon = 0 }
	=
	- P_{ \cal H }
		\Bigl(
			s_\alpha ( q ( \tau ) )^{ -1 }
			\frac{ d s_\alpha ( q ( \tau ) ) }{ d \tau }
		\Bigr).
	\label{25}
\end{equation}
This result may be interpreted that
for a small change of the parameter $ q( \tau ) $ in the base manifold $ Q $,
the lifted point in the fiber space $ G $ moves along the shortest path,
{\it i.e.}, it acquires the smallest change.
Now we notice that the right-hand side of (\ref{25})
is nothing but (the pullback of) the canonical connection $ A $,
which is just the Maurer-Cartan 1-form $ g^{-1} dg $
projected down to the subalgebra $ {\cal H} $,
\begin{equation}
	A = P_{\cal H} ( g^{-1} dg ).
	\label{26}
\end{equation}
It is worth mentioning that
this connection is invariant under the $G$-action over
the homogeneous space $Q$ and provides various topological solutions
of the Yang-Mills equation,
for instance,
the Dirac monopole and the BPST instanton on $Q = S^2$ and $S^4$,
respectively (see, for example \cite{Bais,McMullan}).
\par
Writing the pullback of $ A $ by the section $ s_\alpha $ as
$ A_\alpha = P_{\cal H} ( s_\alpha^{-1} ds_\alpha ) $,
we can write the solution of (\ref{25}) as
$
h_{\alpha\alpha}[q_{k+1}] = h(\epsilon)
= {\cal P}\, \exp [ - \int_{ q_{k} }^{ q_{k+1} } A_{ \alpha } ]
$,
where the symbol ${\cal P}$ denotes the path-ordering.
Using this and gathering scattered pieces, we finally obtain the
propagator (\ref{21}) in the desired form,
\begin{equation}
	K_{ \alpha_n \alpha_0 }^\chi ( q_n , q_0 ; t )
	=
	\int_{q_0}^{q_n} [ dq ] \,
	\rho_\chi ( h_{ \alpha_n \alpha_0 } [ q ] )
	\exp
	\left[ i
		\int dt
		\biggl\{
			\frac{1}{2}
			\biggl| \biggl| \frac{ d q }{ d t } \biggr| \biggr|^2
			- V ( q )
		\biggr\}
	\right],
	\label{27}
\end{equation}
where $ h_{ \alpha_n \alpha_0 } $ is a nonabelian weight factor
\begin{eqnarray}
	&&
        h_{ \alpha_n \alpha_0 } [q]
	\, = \,
	t_{ \alpha_n \alpha_{n-1} } ( q_n )
	\, {\cal P}\,\exp [ - \int_{ q_{n-1} }^{ q_n } A_{ \alpha_{n-1} } ]
	\nonumber
	\\
	&&
        \qquad \times \, t_{ \alpha_{n-1} \alpha_{n-2} } ( q_{n-1} )
	\, {\cal P}\exp [ - \int_{ q_{n-2} }^{ q_{n-1} } A_{ \alpha_{n-2} } ]
	\cdots
	t_{ \alpha_1 \alpha_0 } ( q_1 )
	\, {\cal P}\exp [ - \int_{ q_0 }^{ q_1 } A_{ \alpha_0 } ]
	\label{28}
\end{eqnarray}
with $ q_k \in D_{k-1} \cap D_{k} $
$ ( k = 1 , \cdots , n ) $  being intermediate points.
The factor $ h_{ \alpha_n \alpha_0 }[q] $ is actually a holonomy
associated with the path $ q : [ 0, t ] \to Q $,
and the above expression (\ref{28})
shows that the gauge field interacts minimally in the nonabelian
sense \cite{McMullan}.
We therefore
reached the path integral
(\ref{27}) with (\ref{28}) which precisely reproduces
the result found earlier
in algebraic approaches~\cite{Landsman,LL,Ohnuki}.
%
%
\section{Concluding remarks}
We considered the path integral on a homogeneous space $ Q = G/H $,
and showed that the propagator on $ Q $
can be reduced from the one on $ G $
by integration of redundant degrees of freedom in the fiber
direction of $ H $, with
a non-trivial weight factor $ \rho_\chi (h) $ multiplied.
Being a unitary representation of $ H $, the factor $ \rho_\chi (h) $
preserves the composition law of the propagator, and
a different $ \rho_\chi $ leads to a different (inequivalent)
quantization.
The composition law then allows for a
decomposition of the propagator into small intervals,
and integration over intermediate points eventually
results in the path integral expression.
Examination of the propagator at short distance reveals that
a gauge field is induced in the path integral in the form of
the canonical connection.
Thus we have shown that our guiding principle ---
`first lift and then project' ---
yields the inequivalent quantizations and the induced gauge field correctly.
The basic tool used here is essentially the one
used in \cite{LL}, where the same path integral expression
has been derived from the self-adjoint Hamiltonian through
the Trotter formula.
In this paper we put an emphasis on the role of the geometry
and adopted the guiding principle in order to reach the path integral
expression, rather than defining the quantum theory algebraically first.
\par
Several questions are still left open.
One obvious question is
how we construct a quantum theory on inhomogeneous spaces.
Inhomogeneous spaces often arise in physics,
with the one most frequently discussed being
a Riemann surface with higher genus.
Actually our formulation is not restricted to homogeneous spaces.
A more general situation which allows our principle to be employed
is the following\footnote{%
Such a situation has already been
considered by Montgomery~\cite{Montgomery}
in investigating geometric properties of induced gauge fields
of deformable bodies.}.
Let $ P $ be a Riemannian manifold with a metric $ \tilde{g} $ and
let a Lie group $ H $ act on $ P $ freely and isometrically.
Then the manifold $ M = P/H $ admits an induced metric $ g $,
with which the projection $ \pi : P \to M $
defines a principal bundle and a Riemannian submersion.
Assume that a propagator in $ P $ is $ H $-invariant,
$ K_P ( p'h, ph; t ) = K_P ( p', p; t ) $.
Then our formulation of the path integral can be applied
straightforwardly.
Indeed, the propagator on $M$ can be defined by
\begin{equation}
	K_M^\chi ( p', p; t )
	=
	\int_H dh \, \rho_\chi(h) K_P ( p'h, p; t ),
\end{equation}
which acts on a $ \chi $-equivariant function $ f : P \to V_\chi $;
$ f ( p h ) = \rho_\chi ( h )^{-1} f ( p ) $.
When the base space $ ( M, g ) $ is fixed,
inequivalent quantizations are classified by
choice of the principal bundle $ ( P, M, \pi, H ) $,
the lifted metric $ ( P, \tilde{g} ) $
and the representation $ ( H, V_\chi, \rho_\chi ) $.
However, this scheme may be too general;
we do not have any criterion to choose a specific quantization.
In fact, in this scheme the choice of $ ( P, M, \pi, H, \tilde{g} ) $
is equivalent to introduction of an arbitrary gauge field by hand and,
as a result, we have no longer a natural explanation
of inducing gauge fields.
\par
In contrast, there exists such a criterion
when the base space $ M $ is a homogeneous space $ Q = G/H $.
In fact, the invariance under the $ G $-action determines
both $ g $ and $ \tilde{g} $ uniquely, and hence
the induced gauge field, too.
The only remaining arbitrariness is the choice of the
representation $ \rho_\chi $, and accordingly there are
(infinitely) many inequivalent quantizations.
We may therefore conclude that the existence of
inequivalent quantizations is the norm when quantizing on
a general Riemannian manifold $ ( M, g ) $.
If $ M $ admits a transitive action of some isometry group $ G $,
then the request of invariance will severely restrict possible quantizations.
If $ M $ does not admit such an action,
even a self-adjoint momentum operator cannot be defined globally as
a generator of the transitive action, and hence in that case
we are forced to give up the concept of momentum.
\par
This last point may be important in realizing the significance
of our guiding principle.  Indeed, given a homogeneous space $Q = G/H$
there appears no compelling reason, at a glance,
to lift it to $G$ and consider quantization there.
But this way we can guarantee that there exists a self-adjoint
Hamiltonian given by the quadratic Casimir, which in turn ensures
unitary time evolution of the system.  The existence of such a
Hamiltonian is by no means guaranteed for a system whose configuration
space is nontrivial.  Unfortunately, this is not derived
on the sole ground of geometry,
and finding such a derivation will be crucial in developing
a path integral based on a purely geometric and intuitive principle.
\par
As a final remark,
we add that we have begun a preliminary investigation in two
dimensions into
the meaning of nontrivial topology in field theories
which admit inequivalent quantizations (see, for example~\cite{sigma}).
We have however left untouched the higher dimensional cases,
not to mention the path integral approach.
%
%
\section*{Acknowledgments}
S.T. wishes to thank T.~Iwai and Y.~Uwano for their encouragement
and helpful discussions.
I.T. is grateful to D. McMullan for useful advice.
This work is supported in part by the Grant-in-Aid for Scientific
Research from the Ministry of Education, Science and Culture (No. 07804015).
%
%

%
\end{document}